# Outburst characteristics of the dwarf nova SDSS J073208.11+413008.7

Jeremy Shears, Robert Koff, Richard Sabo, Bart Staels, William Stein and Patrick Wils

## Abstract

We report unfiltered photometry during the first confirmed superoutburst of the recently discovered dwarf nova, SDSS J073208.11+413008.7 and conclude that it is a member of the SU UMa family. At its brightest, the star was magnitude 15.7. The outburst amplitude was nearly 5 magnitudes and it lasted about 14 days. We determined the mean superhump period from our first 3 nights of observations as $P_{sh}$ = 0.07979(19) d, however analysis of the O-C residuals showed a dramatic evolution in $P_{sh}$ during the outburst. During the first part of the plateau phase the period increased with $dP_{sh}/dt$ = +2.81(9) x $10^{-3}$. There was then an abrupt change following which the period decreased with $dP_{sh}/dt$ = -0.78(12)×$10^{-3}$. The amplitude of the superhumps also varied, with a maximum amplitude near the beginning of the outburst and a second maximum later in the plateau phase. Analysis of archival data suggests that the star undergoes frequent outbursts: we identified 10 normal outbursts and at least 3 likely superoutbursts over a 4.5 year interval. We estimate that the superoutburst period is around 215 days.

## Introduction

Dwarf novae are a class of cataclysmic variable star in which a white dwarf primary accretes material from a secondary star via Roche lobe overflow. The secondary is usually a late-type main-sequence star. In the absence of a significant white dwarf magnetic field, material from the secondary passes through an accretion disc before settling on the surface of the white dwarf. As material builds up in the disc, a thermal instability is triggered that drives the disc into a hotter, brighter state causing an outburst in which the star apparently brightens by several magnitudes [1]. Dwarf novae of the SU UMa family occasionally exhibit superoutbursts which last several times longer than normal outbursts and may be up to a magnitude brighter. During a superoutburst the light curve of a SU UMa system is characterised by superhumps. These are modulations in the light curve with a period a few percent longer than the orbital period. They are thought to arise from the interaction of the secondary star orbit with a slowly precessing eccentric accretion disc. The eccentricity of the disc arises because a 3:1 resonance occurs between the secondary star orbit and the motion of matter in the outer accretion disc. For a more detailed review of SU UMa dwarf novae and superhumps, the reader is directed to reference 1.

SDSS J073208.11+413008.7 was found in the course of a variability search for dwarf novae [2] in data from the Sloan Digital Sky Survey (SDSS; Data Release 7 [3]). SDSS lists the object as having g = 20.7 and r = 20.4.

The outburst discussed in the paper was detected by Shears [4] on 2009 Dec 31.852 at an unfiltered CCD magnitude of 16.1 during a real-time patrol for outbursts of dwarf novae.

**Photometry and analysis**

During the outburst, the authors conducted 138 hours of unfiltered time-resolved photometry using the instrumentation shown in Table 1 and according to the observation log in Table 2. Images were dark-subtracted and flat-fielded prior to being measured using differential aperture photometry relative to either GSC 2966-1413 (V=14.30) or GSC 2966-1521 (V=14.39); magnitudes from GSC 2.2. Given that each observer used slightly different instrumentation, including CCD cameras with different spectral responses, small systematic differences are likely to exist between observers. Where overlapping datasets were obtained, we aligned measurements by different observers by experiment. Adjustments of up to 0.08 magnitude were made. However, given that the aim of the time resolved photometry was to investigate periodic variations in the light curve, we consider this not to be a significant disadvantage. Heliocentric corrections were applied to all data.

**Outburst light curve**

The overall light curve of the outburst is shown in the top panel of Figure 1, based on the authors' photometry. The star appeared to have brightened between detection and the next observation 2 nights later. Thus the detection was probably made during the rise to maximum. The star was observed to be at its brightest at magnitude 15.7 on JD 2455199. Taking the SDSS magnitude of r = 20.4 as quiescence (our unfiltered CCDs are slightly red sensitive, so the r band measurement is likely a better comparison than g), the outburst amplitude is nearly 5 magnitudes.

Time series photometry commenced on JD 2455199, some 2 days after the outburst detection. The 10 days from JD 2445199 and 2455209 corresponds to the plateau phase during which the star faded gradually at a mean rate of 0.09 mag/d. There then followed a rapid decline (0.68 mag/d) over the next 2 days reaching mag ~18.0 after which no further observations were made. Thus the outburst lasted about 14 days.

**Measurement of the superhump period**

In Figure 2 we plot expanded views of the longer time series photometry runs, where each panel shows 2 days of data drawn to the same scale. This clearly shows the presence of regular modulations which we interpret as superhumps. The presence of superhumps is diagnostic that SDSS J073208.11+413008.7 is a member of the SU UMa family of dwarf novae, making this the first confirmed superoutburst of the star.

To study the superhump behaviour, we first extracted the times of each sufficiently well-defined superhump maximum by fitting a quadratic function to the individual light curves. Times of 58 superhump maxima were found and are listed in Table 3. An analysis of the times of maximum for cycles 0 to 27 (JD 2455199 to 2455201) assuming a linear fit allowed us to obtain the following linear superhump maximum ephemeris:

$$HJD_{max} = 2455199.70539(24) + 0.07979(19) \times E \qquad \text{Equation 1}$$

This gives the mean superhump period for the first three nights of the superoutburst $P_{sh}$ = 0.07979(19) d. The O–C residuals for the superhump maxima for the complete outburst relative to the ephemeris are shown in the middle panel in Figure 1.

**Superhump evolution**

The O-C diagram shows that the superhump period changed significantly during the outburst. Kato *et al.* [5] studied the superhump period evolution in a large number of SU UMa systems and found that many outbursts appeared to show three distinct stages: an early evolutionary stage (A) with longer superhump period, a middle stage (B) during which systems with $P_{orb}$ < 0.08 d have a positive period derivative, and a final stage (C) with a shorter $P_{sh}$. Our O-C diagram for SDSS J073208.11+413008.7 is consistent with this interpretation although it appears we missed stage A. The interval between JD 2455199 and 2455202 (blue data points in Figure 1, middle panel) corresponds to Stage B, during which we found an increase in the superhump period with $dP_{sh}/dt$ = +2.81(9) x $10^{-3}$ by fitting a quadratic function to the data (blue dotted line in Figure 1). There was then a change in period at around JD 2455203, following which the period decreased with $dP_{sh}/dt$ = -0.78(12)×$10^{-3}$ during stage C as shown by the quadratic fit to the data between JD 2455204 and JD 2455209 (green dotted line in Figure 1). The period transition from stage B to C was sudden, as is common in SU UMa systems including SW UMa and UV Per [5, 6] and ASAS J224349+0809.5 [7]. Close inspection of the outburst light curve suggests that the transition in superhump regime appeared to correspond with a temporary slowing in the fading trend during the plateau phase.

We also found that the superhump peak-to-peak amplitude changed significantly during the outburst (Figure 1, bottom panel). During the first night of observation (JD 2455199), the superhump amplitude was 0.20 magnitude, after which it gradually decreased to 0.13 magnitude on JD 2455202. The superhumps then increased in amplitude, reaching a second maximum of 0.20 magnitudes between JD 2455205 and 2455207. The amplitude decreased rapidly during the rest of the outburst; the last observed superhumps had an amplitude of 0.07 magnitudes. The second maximum occurred during the second half of the plateau phase and was well after the transition from stage B to C. This is in contrast to what was observed with SW UMa [5, 6] and ASAS J224349+0809.5 [7] where the second superhump amplitude maximum coincided with the discontinuity between stage B and C.

**Orbital period**

We searched for the orbital signal in the power spectrum of the photometric data using a variety of statistical techniques. We clearly observed signals corresponding to the superhump period and its aliases, but no convincing signal which we could attribute to $P_{orb}$ was forthcoming (data not shown).

Gaensicke et al. [8] developed an empirical relation between $P_{orb}$ and $P_{sh}$ by analysing a wide range of SU UMa systems:

$$P_{orb} = 0.9192(52)\ P_{sh} + 5.39(52) \qquad \text{Equation 2}$$

where both periods are given in minutes. Using our value of $P_{sh} = 0.07979(19)$ d in this equation, allows us to estimate $P_{orb} = 0.07708(96)$ d.

**Outburst frequency**

We examined data from the Catalina Real-Time Transient Survey [9] (CRTS) for outbursts of SDSS J073208.11+413008.7. CRTS observed the field on 47 nights spanning some 4.5 years and the object was in outburst on 14 nights. This represents about 30% of the time, suggesting that it is a very active dwarf nova. Taking outbursts reaching magnitude 16.5 or fainter as being normal outbursts and those brighter than magnitude 16 as superoutbursts, we find 10 such normal outbursts and 3 superoutbursts (one superoutburst was observed on 2 nights). The superoutbursts were observed in 2006 February, 2008 April and 2008 December. Of course one or more of the "normal" outbursts could have been superoutbursts if the brightest part of the outburst was missed. Furthermore other outbursts and superoutbursts may have been missed completely due to the incomplete coverage by CRTS.

On one occasion, two normal outbursts were separated by 20 days and on another occasion a normal outburst occurred only 30 days after a superoutburst. In both instances there is a "fainter than" observation between the outbursts, which means they are indeed separate outbursts. We also note that following the superoutburst reported in this paper, another normal outburst was detected only 15 days later on 2010 Jan 29 (JD 2455226) at magnitude 17.0 and a further one 24 days after that, on Feb 22 (JD 2455250) at magnitude 17.4. These observations further indicate that the normal outburst cycle is very short, possibly of the order of 2 to 4 weeks.

We have adapted a method introduced by Southworth et al. [10] for normal outbursts in order to estimate the length of the superoutburst cycle for our SU UMa system, based on the typical superoutburst duration, the observed number of outbursts and the actual times of observation. Using the CRTS data, our observation that a superoutburst lasts 14 days, and assuming that the superoutbursts are quasi-periodic (i.e. with the length of the superoutburst cycle following a Gaussian distribution, with a standard deviation of a quarter of its mean), we performed Monte

Carlo simulations which suggest that there is a 95% or higher probability of CRTS observing 3 superoutbursts when the supercycle is between 105 and 550 days. The highest probability is at 215 days and we note that the time between the 2008 April and December superoutbursts is 265 days, which is consistent with the statistics. However, given that it is possible that some of the normal outbursts may have been superoutbursts, we plot the probability of observing 3, 4 or 5 outbursts for a given length of superoutburst period as Figure 3. This shows that the highest probability of observing 3, 4, or 5 superoutbursts is at 215, 160 and 134 days respectively. Thus the supercycle may actually be less than 215 days. Only further observational coverage will confirm this.

**Conclusions**

We report unfiltered photometry during the first confirmed superoutburst of the recently discovered dwarf nova, SDSS J073208.11+413008.7. The presence of superhumps shows that it is a member of the SU UMa family. The outburst amplitude was nearly 5 magnitudes and it lasted about 14 days with a maximum brightness of magnitude 15.7.

We determined the mean superhump period from our first 3 nights of observations as $P_{sh}$ = 0.07979(19) d, however analysis of the O-C residuals showed a dramatic evolution in $P_{sh}$ during the outburst. During the first part of the plateau phase the period increased with $dP_{sh}/dt$ = +2.81(9) x $10^{-3}$. There was then an abrupt change following which the period decreased with $dP_{sh}/dt$ = --0.78(12)×$10^{-3}$. The amplitude of the superhumps also varied, with a maximum amplitude near the beginning of the outburst and a second maximum later in the plateau phase.

Using an empirical relationship between $P_{sh}$ and $P_{orb}$ established in a variety of SU UMa systems, we estimate the orbital period of SDSS J073208.11+413008.7 as $P_{orb}$ = 0.07708(96).

Analysis of archival data suggests that the star undergoes frequent outbursts: we identified 10 normal outbursts and at least 3 likely superoutbursts. Assuming that the superoutbursts are periodic, we estimate that the outburst period is around 215 days.

**Acknowledgements**

The authors gratefully acknowledge the use of data from the Catalina Real-Time Transient Survey [9]. We also used SIMBAD and Vizier, operated through the Centre de Donées Astronomiques (Strasbourg, France), and the NASA/Smithsonian Astrophysics Data System. We thank the referees, Dr Chris Lloyd and Dr Robert Connon Smith for helpful comments which have improved the paper. JS thanks the Council of the British Astronomical Association for the award of a Ridley Grant that was used to purchase some of the equipment used in this research.

**Addresses**

JS: "Pemberton", School Lane, Bunbury, Tarporley, Cheshire, CW6 9NR, UK [bunburyobservatory@hotmail.com]
RK: 980 Antelope Drive West, Bennett, CO 80102, USA [bob@AntelopeHillsObservatory.org]
RS: 2336 Trailcrest Dr., Bozeman, MT 59718, USA [richard@theglobal.net]
BS: CBA Flanders, Patrick Mergan Observatory, Koningshofbaan 51, Hofstade, Aalst, Belgium [staels.bart.bvba@pandora.be]
WS: 6025 Calle Paraiso, Las Cruces, NM 88012, USA [starman@tbelc.org]
PW: Vereniging voor Sterrenkunde, Belgium [patrickwils@yahoo.com]

| Observer | Telescope | CCD |
|---|---|---|
| Koff | 0.25 m SCT | Apogee AP-47 |
| Sabo | 0.43 m reflector | SBIG STL-1001 |
| Shears | 0.28 m SCT | Starlight Xpress SXVF-H9 |
| Staels | 0.28 m SCT | Starlight Xpress MX-716 |
| Stein | 0.35 m SCT | SBIG ST10XME |

**Table 1: Equipment used**

| Start time | End time | Duration (h) | Observer |
|---|---|---|---|
| 2455199.620 | 2455199.991 | 8.9 | Stein |
| 2455200.407 | 2455200.465 | 1.4 | Shears |
| 2455200.442 | 2455200.582 | 3.4 | Staels |
| 2455200.641 | 2455200.734 | 2.2 | Sabo |
| 2455200.665 | 2455201.055 | 9.4 | Stein |
| 2455201.386 | 2455201.414 | 0.7 | Shears |
| 2455201.652 | 2455202.052 | 9.6 | Stein |
| 2455202.637 | 2455203.017 | 9.1 | Stein |
| 2455203.579 | 2455203.766 | 4.5 | Sabo |
| 2455204.285 | 2455204.666 | 9.1 | Staels |
| 2455204.291 | 2455204.470 | 4.3 | Shears |
| 2455204.576 | 2455204.821 | 5.9 | Sabo |
| 2455205.582 | 2455205.970 | 9.3 | Stein |
| 2455206.739 | 2455207.045 | 7.3 | Stein |
| 2455207.588 | 2455208.034 | 10.7 | Koff |
| 2455207.597 | 2455207.972 | 9.0 | Sabo |
| 2455208.573 | 2455208.777 | 4.9 | Koff |
| 2455208.591 | 2455209.010 | 10.1 | Sabo |
| 2455208.620 | 2455209.049 | 10.3 | Stein |
| 2455209.622 | 2455209.963 | 8.2 | Stein |
| 2455211.620 | 2455211.640 | 0.5 | Sabo |

**Table 2 : Log of time-series photometry**

| Superhump cycle | Superhump maximum (HJD) | O-C (d) | Error (d) | Superhump amplitude |
|---|---|---|---|---|
| 0 | 2455199.7033 | -0.0021 | 0.0005 | 0.18 |
| 1 | 2455199.7853 | 0.0002 | 0.0006 | 0.19 |
| 2 | 2455199.8660 | 0.0010 | 0.0013 | 0.17 |
| 3 | 2455199.9474 | 0.0027 | 0.0014 | 0.20 |
| 13 | 2455200.7433 | 0.0007 | 0.0011 | 0.16 |
| 14 | 2455200.8205 | -0.0019 | 0.0004 | 0.18 |
| 15 | 2455200.8999 | -0.0023 | 0.0014 | 0.16 |
| 16 | 2455200.9823 | 0.0003 | 0.0015 | 0.18 |
| 25 | 2455201.7023 | 0.0022 | 0.0008 | 0.15 |
| 26 | 2455201.7792 | -0.0008 | 0.0009 | 0.16 |
| 27 | 2455201.8599 | 0.0002 | 0.0010 | 0.14 |
| 28 | 2455201.9390 | -0.0005 | 0.0013 | 0.17 |
| 29 | 2455202.0211 | 0.0018 | 0.0020 | 0.18 |
| 37 | 2455202.6643 | 0.0067 | 0.0007 | 0.15 |
| 38 | 2455202.7420 | 0.0046 | 0.0005 | 0.13 |
| 39 | 2455202.8226 | 0.0054 | 0.0007 | 0.15 |
| 40 | 2455202.9046 | 0.0076 | 0.0018 | 0.16 |
| 41 | 2455202.9804 | 0.0036 | 0.0020 | 0.16 |
| 49 | 2455203.6140 | -0.0011 | 0.0024 | 0.16 |
| 50 | 2455203.6983 | 0.0034 | 0.0023 | 0.16 |
| 58 | 2455204.3384 | 0.0052 | 0.0021 | ND |
| 59 | 2455204.4190 | 0.0060 | 0.0019 | 0.16 |
| 59 | 2455204.4193 | 0.0063 | 0.0020 | 0.17 |
| 62 | 2455204.6559 | 0.0036 | 0.0018 | 0.17 |
| 63 | 2455204.7370 | 0.0048 | 0.0015 | 0.16 |
| 64 | 2455204.8119 | 0.0000 | 0.0015 | 0.18 |
| 74 | 2455205.6121 | 0.0023 | 0.0021 | 0.19 |
| 75 | 2455205.6921 | 0.0025 | 0.0007 | 0.19 |
| 76 | 2455205.7702 | 0.0008 | 0.0006 | 0.20 |
| 77 | 2455205.8446 | -0.0046 | 0.0007 | 0.19 |
| 78 | 2455205.9215 | -0.0075 | 0.0017 | 0.17 |
| 89 | 2455206.8032 | -0.0035 | 0.0006 | 0.17 |
| 90 | 2455206.8794 | -0.0071 | 0.0005 | 0.16 |
| 91 | 2455206.9590 | -0.0073 | 0.0009 | 0.20 |
| 100 | 2455207.6716 | -0.0128 | 0.0010 | 0.17 |
| 100 | 2455207.6771 | -0.0073 | 0.0017 | 0.17 |
| 101 | 2455207.7542 | -0.0099 | 0.0013 | 0.17 |
| 101 | 2455207.7464 | -0.0178 | 0.0023 | 0.16 |
| 102 | 2455207.8341 | -0.0099 | 0.0032 | 0.20 |
| 102 | 2455207.8285 | -0.0155 | 0.0025 | 0.19 |
| 103 | 2455207.9102 | -0.0136 | 0.0036 | 0.18 |
| 103 | 2455207.9132 | -0.0106 | 0.0038 | 0.19 |
| 104 | 2455207.9889 | -0.0147 | 0.0039 | 0.19 |
| 112 | 2455208.6231 | -0.0188 | 0.0021 | 0.15 |
| 113 | 2455208.7029 | -0.0188 | 0.0020 | 0.13 |
| 113 | 2455208.7006 | -0.0211 | 0.0005 | 0.14 |
| 113 | 2455208.7008 | -0.0209 | 0.0021 | 0.13 |
| 114 | 2455208.7804 | -0.0210 | 0.0012 | 0.12 |
| 114 | 2455208.7808 | -0.0207 | 0.0012 | 0.13 |
| 115 | 2455208.8623 | -0.0190 | 0.0007 | 0.11 |
| 115 | 2455208.8639 | -0.0173 | 0.0014 | 0.12 |
| 116 | 2455208.9403 | -0.0208 | 0.0009 | 0.12 |
| 116 | 2455208.9419 | -0.0191 | 0.0024 | 0.12 |
| 117 | 2455209.0210 | -0.0198 | 0.0003 | 0.11 |
| 125 | 2455209.6529 | -0.0262 | 0.0021 | 0.10 |
| 126 | 2455209.7358 | -0.0232 | 0.0020 | 0.09 |
| 127 | 2455209.8086 | -0.0301 | 0.0029 | 0.07 |
| 128 | 2455209.8903 | -0.0282 | 0.0026 | 0.07 |

**Table 3: Superhump maximum times and amplitudes**
ND: not determined

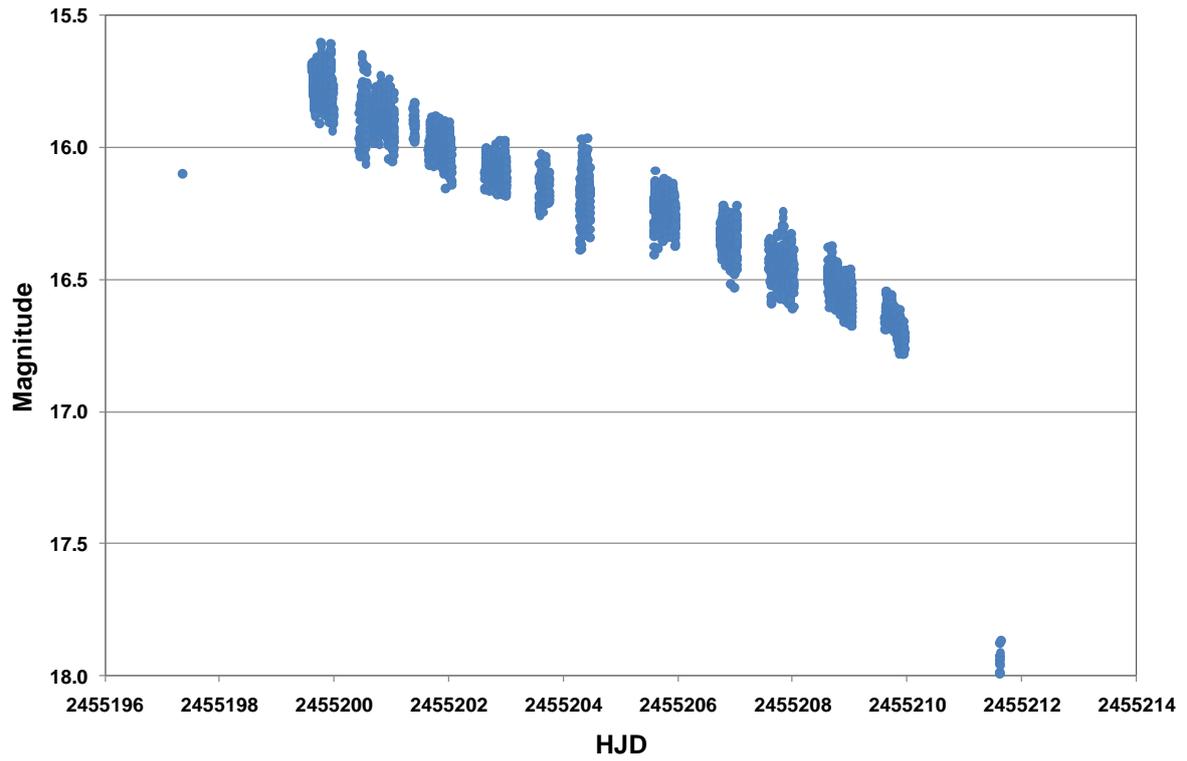
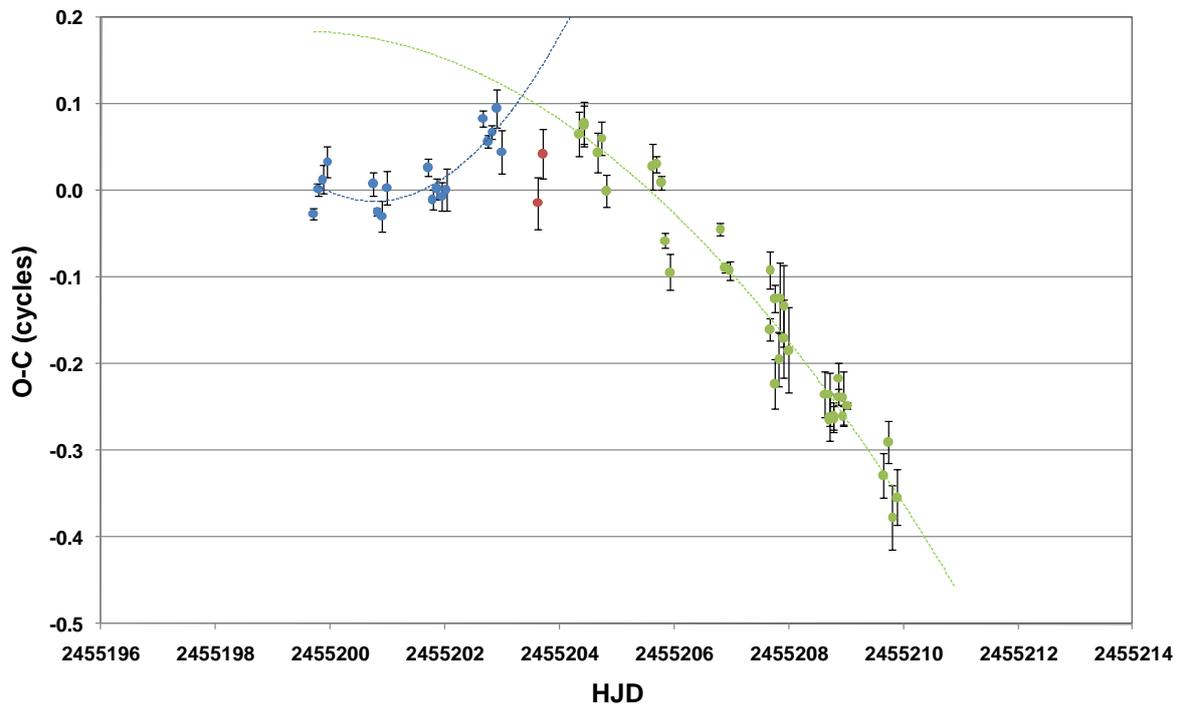

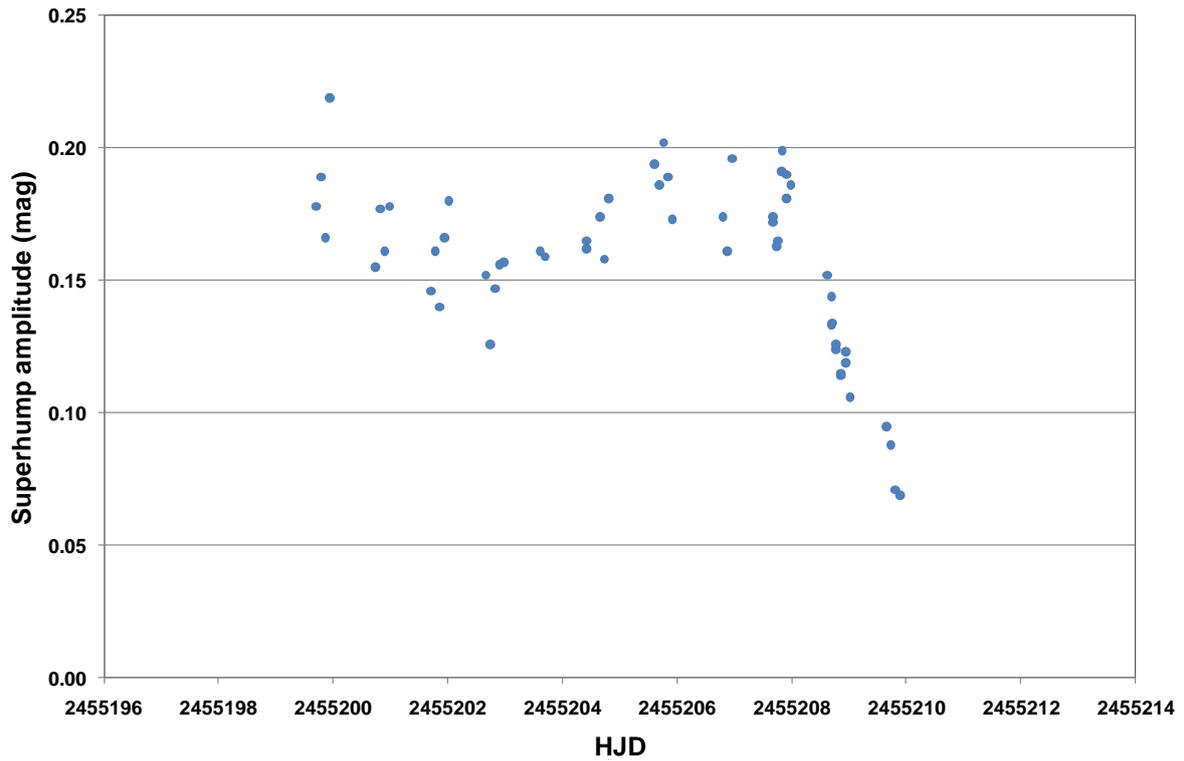

**Figure 1: Light curve of the outburst (top), O-C diagram of superhump maxima relative to the ephemeris in Equation 1 (middle) and superhump amplitude (bottom)**

In the O-C diagram, the blue dotted line is a quadratic fit to the data between JD 2455199 and 2455202 (blue data points) and the green dotted line is a quadratic fit to the data between JD 2445204 to 2445209 (green data points)

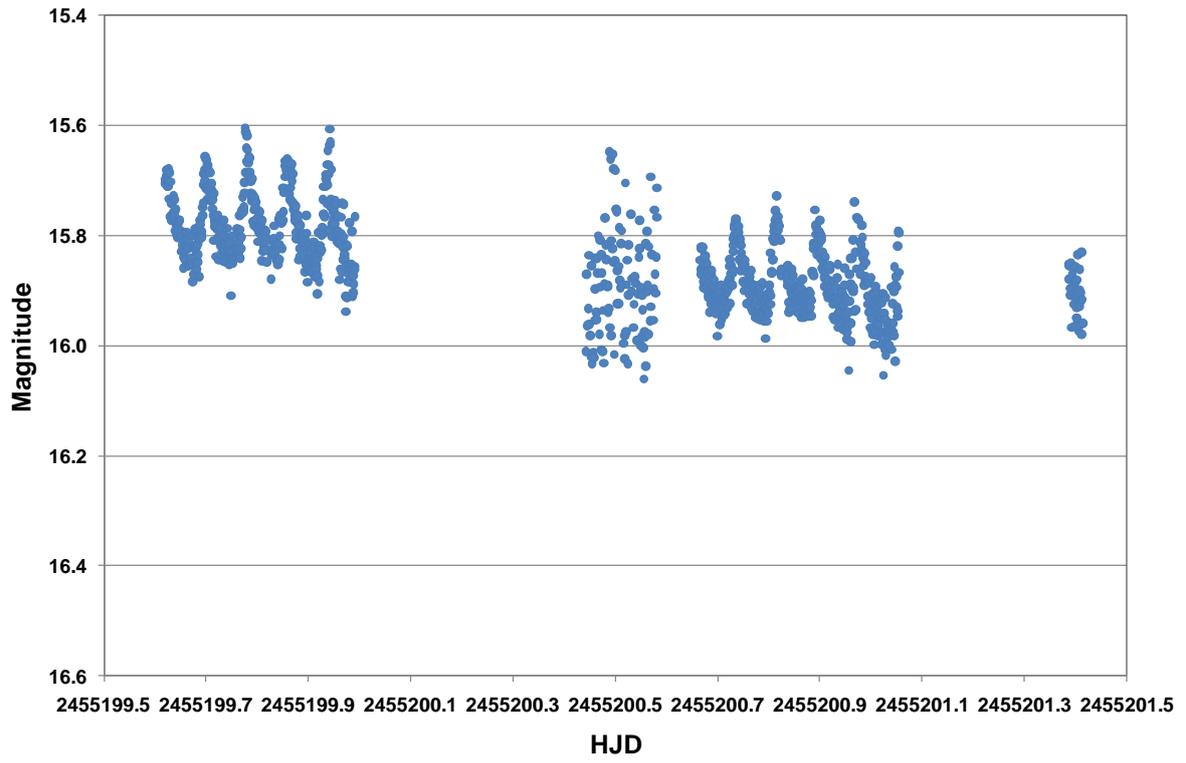

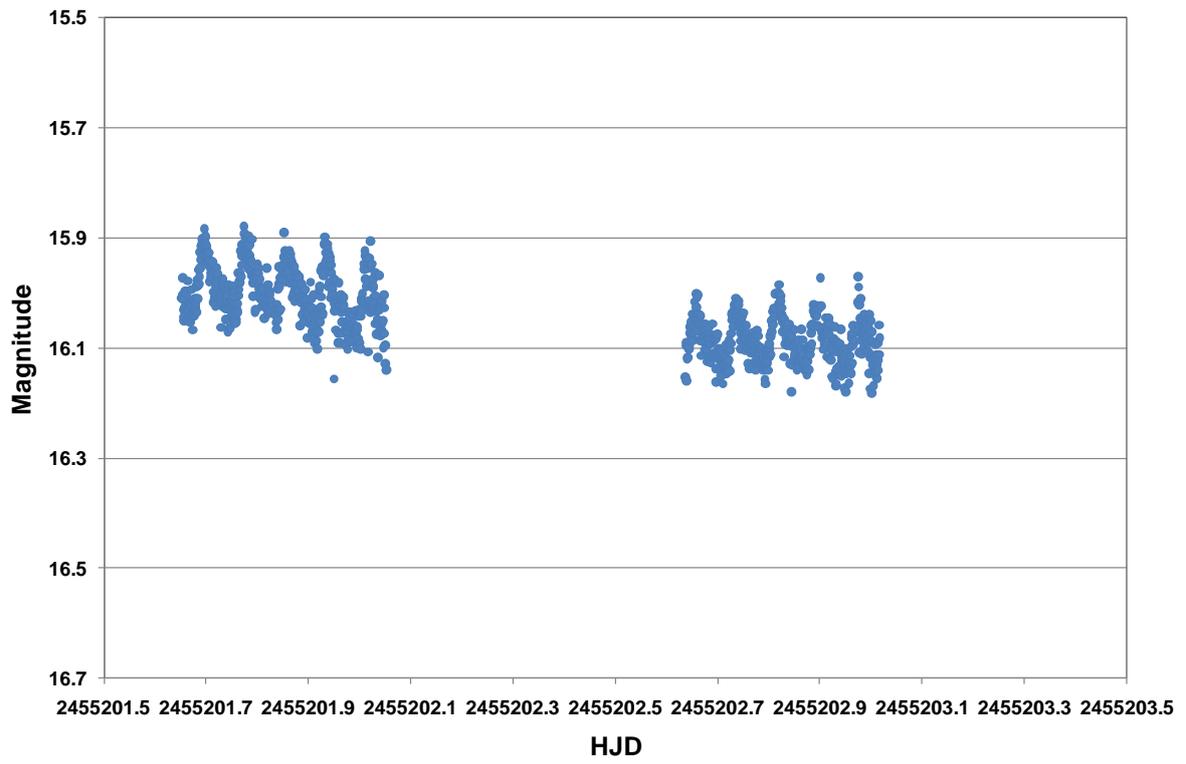

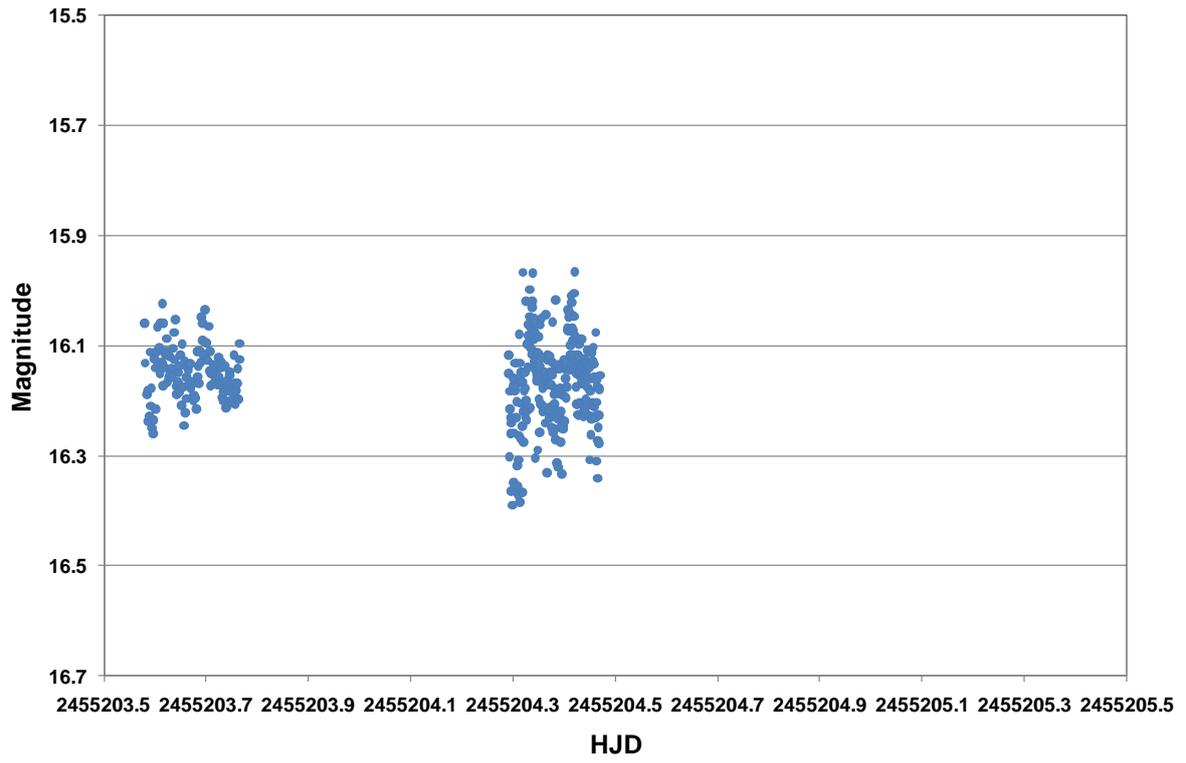

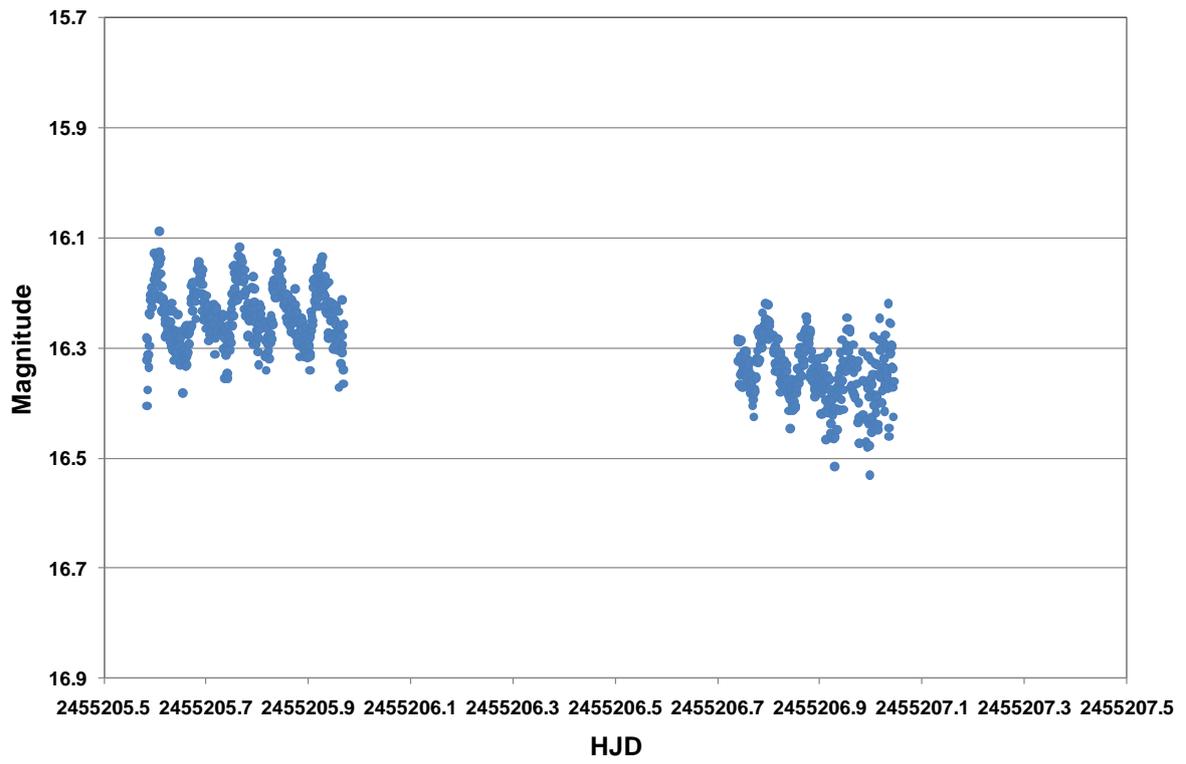

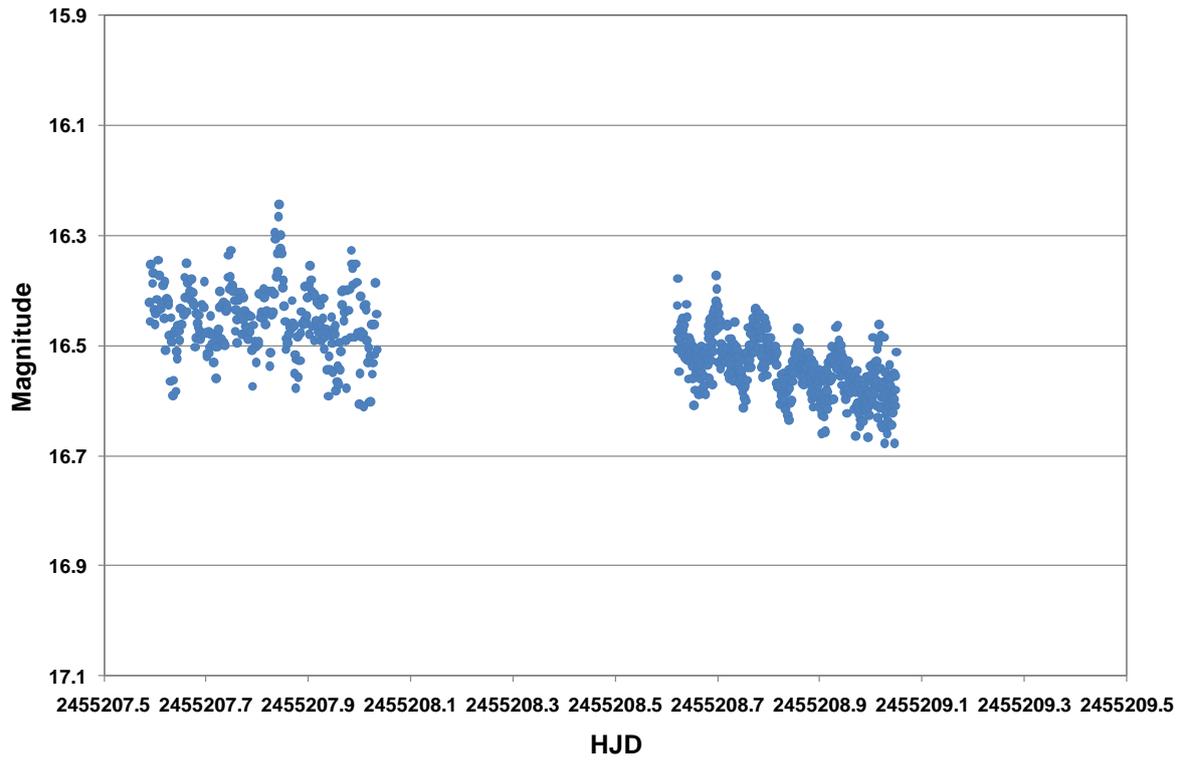

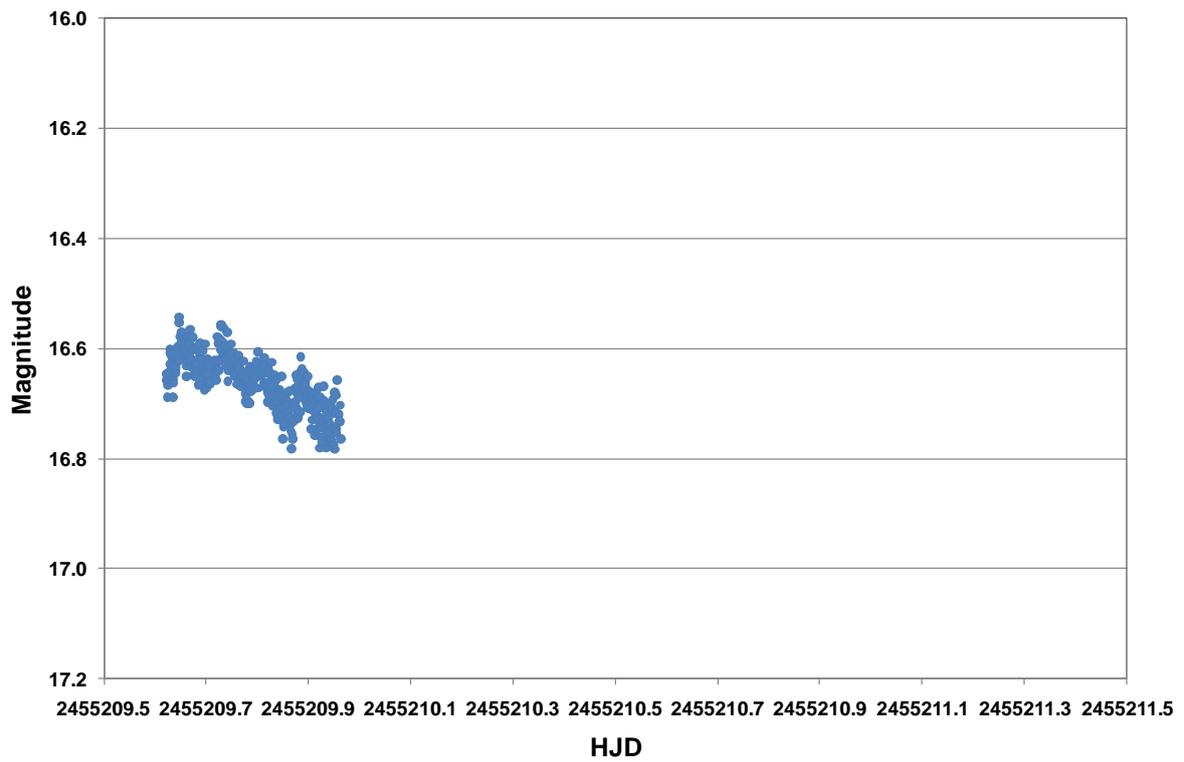

**Figure 2: Time series photometry during the outburst of SDSS J073208.11+413008.7**

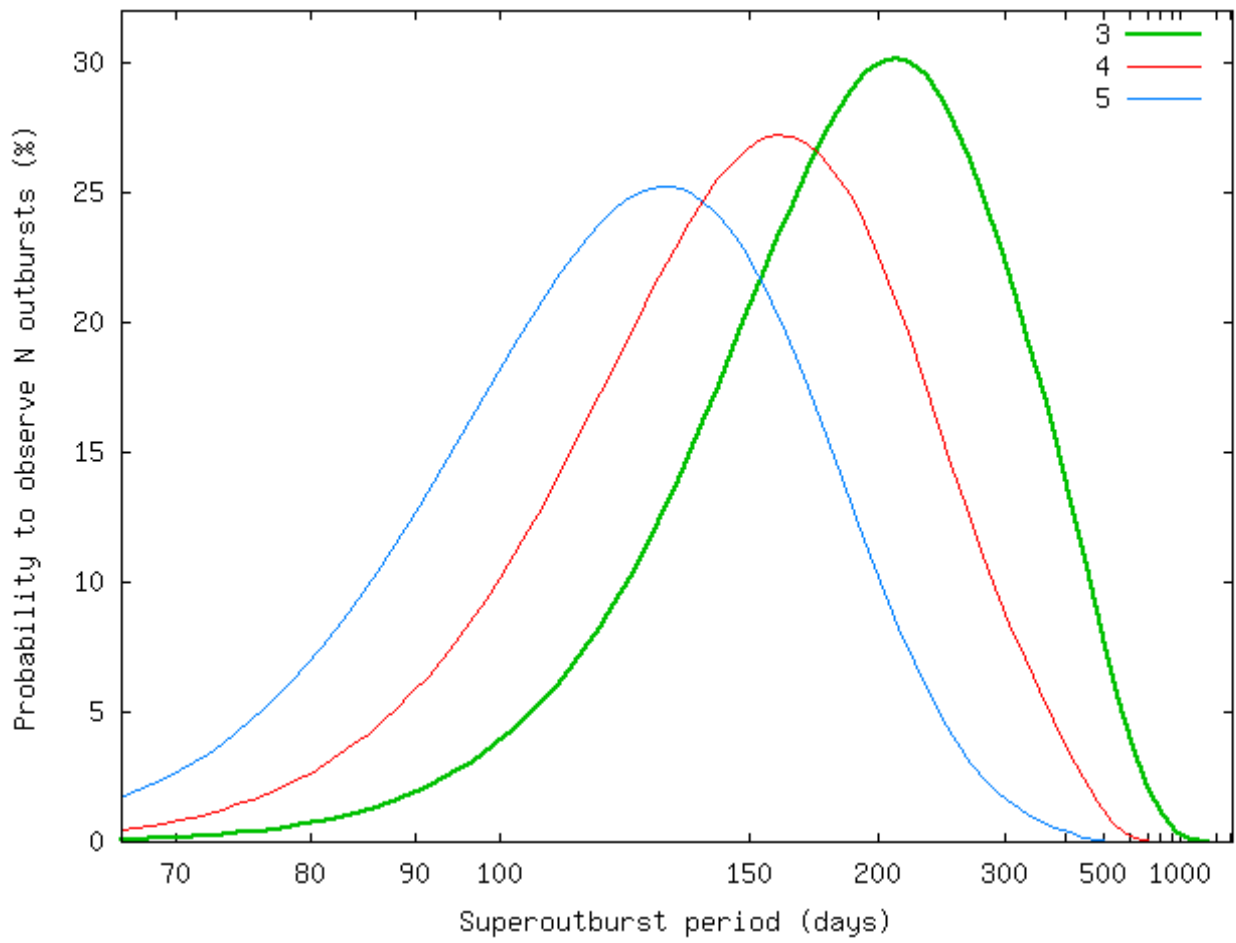

**Figure 3: Probability of observing 3, 4 or 5 superoutbursts for a given length of superoutburst period**